\documentclass[sigconf]{acmart}
\renewcommand\footnotetextcopyrightpermission[1]{} 

\usepackage{booktabs} 
\usepackage{tabularx} 
\usepackage{amsmath}
\usepackage{amssymb}
\usepackage{algorithm}
\usepackage{multirow}
\usepackage{graphicx}
\usepackage{subcaption}
\usepackage{epstopdf}
\usepackage[noend]{algpseudocode}
\usepackage{balance}

\newcommand{\dquotes}[1]{``#1''}

\usepackage{xspace}

\makeatletter
\def\BState{\State\hskip-\ALG@thistlm}
\makeatother

\newcolumntype{L}[1]{>{\raggedright\let\newline\\\arraybackslash\hspace{0pt}}m{#1}}
\newcolumntype{C}[1]{>{\centering\let\newline\\\arraybackslash\hspace{0pt}}m{#1}}
\newcolumntype{R}[1]{>{\raggedleft\let\newline\\\arraybackslash\hspace{0pt}}m{#1}}

\usepackage[moderate]{savetrees}

\newcommand*\diff{\mathop{}\!\mathrm{d}}

\usepackage{pifont}
\newcommand{\cmark}{\ding{51}}%
\setcopyright{none}

\newif\ifmakeshorter
\makeshortertrue

\begin{document}




\title[Recommender Systems Fairness Evaluation via Generalized Cross Entropy]{Recommender Systems Fairness Evaluation\\ via Generalized Cross Entropy}
\titlenote{Permission to make digital or hard copies of part or all of this work for personal or classroom use is granted without fee provided that copies are not made or distributed for profit or commercial advantage and that copies bear this notice and the full citation on the first page. Copyrights for third-party components of this work must be honored \\ For all other uses, contact the owner/author(s). \\ RMSE workshop, The 13th ACM Conference on Recommender Systems (RecSys), 2019,  Copenhagen, Denmark.}
 
\author{Yashar Deldjoo}
\affiliation{%
  \institution{Polytechnic University of Bari, Italy}
}
\email{yashar.deldjoo@poliba.it}

\author{Vito Walter Anelli}
\affiliation{%
  \institution{Polytechnic University of Bari, Italy}
}
\email{vitowalter.anelli@poliba.it}

\author{Hamed Zamani}
\affiliation{%
  \institution{University of Massachusetts}
  \city{Amherst}
  \country{USA}
}
\email{zamani@umass.edu}

\author{Alejandro Bellog\'{i}n}
\affiliation{%
  \institution{Autonomous University of Madrid, Spain}
}
\email{alejandro.bellogin@uam.es}

\author{Tommaso Di Noia}
\affiliation{%
  \institution{Polytechnic University of Bari, Italy}
}
\email{tommaso.dinoia@poliba.it}

\begin{abstract}
Fairness in recommender systems has been considered with respect to sensitive attributes of users (e.g., gender, race) or items (e.g., revenue in a multistakeholder setting). Regardless, the concept has been commonly interpreted as some form of \textit{equality} -- i.e., the degree to which the system is meeting the information needs of all its users \textit{in an equal sense}. In this paper, we argue that fairness in recommender systems does not necessarily imply equality, but instead it should consider a distribution of resources based on merits and needs.
We present a probabilistic framework based on generalized cross entropy to evaluate fairness of recommender systems under this perspective, 
where we show that the proposed framework is flexible and explanatory by allowing to incorporate domain knowledge (through an ideal fair distribution) that can help to understand which item or user aspects a recommendation algorithm is over- or under-representing. Results on two real-world datasets show the merits of the proposed evaluation framework both in terms of user and item fairness. 
\end{abstract}

\ifmakeshorter
\fi





\maketitle

\vspace{-4pt}
\section{Introduction and Context}

Recommender systems (RS) are widely applied across the modern Internet, in e-commerce websites, movies and music streaming platforms, or on social media to point users to items (products or services)~\cite{ekstrand2018all}. For evaluation of RS, accuracy metrics are typically employed, which measure how much the presented items will be of interest to the target user. One commonly raised concern is how much the recommendations produced by RS are fair. For example, do users
of certain gender or race receive fair utility (i.e., benefit) from the recommendation service? To answer this question, one has to recognize the multiple stakeholders involved in such systems and that fairness issues can be studied for more than one group of participants~\cite{burke2017multisided}.
In a job recommendation scenario, for instance, these multiple groups can be the job seekers and prospective employers where fairness toward both parties has to be recognized. 
Moreover, fairness in RS can be measured towards items or users;
in this context, user and item fairness are commonly associated with an equal chance for appearing in the recommendation results (items) or receiving results of the same quality (users).
As an example for the latter, an unfair system may discriminate against users of a particular race or gender. 

One common characteristic of the previous literature focusing on RS fairness evaluation is that fairness has been commonly interpreted as some form of \textit{equality} across multiple groups (e.g., gender, race). For example,~\citet{ekstrand2018all} studied whether RS produce \textit{equal utility} for users of different demographic groups. The authors find demographic differences in measured effectiveness across two datasets from different domains. 
\citet{yao2017beyond} studied various types of unfairness that can occur in collaborative filtering models where, to produce fair recommendations, the authors proposed to penalize algorithms producing disparate distributions of prediction error. For additional resources see~\cite{burke2017multisided,DBLP:journals/corr/abs-1809-09030,DBLP:conf/fat/BurkeSO18,DBLP:conf/cikm/ZhuHC18,DBLP:conf/um/ZhengDMK18}. 

Nonetheless, although less common, there are a few works where fairness has been defined beyond uniformity. 
For instance, in~\cite{DBLP:conf/sigir/BiegaGW18}, the authors proposed an approach focused on mining the relation between \textit{relevance} and \textit{attention} in Information Retrieval by exploiting the positional bias of search results. 
That work promotes the notion that ranked subjects should receive attention that is proportional to their \textit{worthiness} in a given search scenario and achieve fairness of attention by making exposure proportional to relevance.
Similarly, a framework formulation of fairness constraints is presented in \cite{DBLP:conf/kdd/SinghJ18} on rankings in terms of exposure allocation, both with respect to group fairness constraints and individuals. 
Another approach where non-uniform fairness has been used is the work proposed in \cite{DBLP:conf/cikm/ZehlikeB0HMB17}, where the authors aim to solve the top-k ranking problem by optimizing a fair utility function under two conditions: in-group monotonicity (i.e., rank more relevant items above less relevant within the group) and group fairness (proportion of protected group items in the top-k ranking should be above a minimum threshold). 
In summary, even though these approaches use some notion of non-uniformity, they are applied under different perspectives and purposes.


In the present work, we argue that fairness does not necessarily imply equality between groups, but instead proper distribution of utility (benefits)  based on merits and needs. To this end, we present a probabilistic framework for evaluating RS fairness based on attributes of any nature (e.g., sensitive or insensitive) for both items or users and show that the proposed framework is flexible enough to measure fairness in RS by considering fairness as equality or non-equality among groups, as specified by the system designer or any other parties involved in multistakeholder setting. 
As we shall see later, the discussed approaches are different from our proposal in that we are able to accommodate different notions of fairness, not only \textit{ranking}, e.g., \textit{rating}, \textit{ranking} and \textit{even-beyond accuracy metrics}.
In fact, the main advantage of our framework is to provide the system designer with a high degree of flexibility on defining fairness from multiple viewpoints.
%
Results on two real-world datasets show the merits of the proposed evaluation framework, both in terms of user and item fairness.

\begin{table*}[tb]
\caption{A set of 6 users belonging to groups $g_1$ and $g_2$ and 10 items along with their true labels marked by \cmark and recommended items by recommenders Rec 0, Rec 1, Rec 2.
Rec 0 produces 3 and 6 relevant items for free and premium users (in total) respectively; Rec 1 generates 1 relevant item for each user; Rec 2 produces recommended items that are all relevant for all users.\label{tbl:toy_example}}
\vspace{-10pt}
\begin{tabular}{|l|c|c|c|c|c|c|c|c|c|c|c|c|c|c|}
\hline 
  &  &$i_1$ &$i_2$ &$i_3$  &$i_4$ &$i_5$ &$i_6$  &$i_7$ &$i_8$ &$i_9$  &$i_{10}$ &\textbf{Rec 0}  &\textbf{Rec 1}   &\textbf{Rec 2}  \\ \hline
 ${\color[HTML]{FE0000} a_1}$  &{\color[HTML]{FE0000} \textbf{user 1}}  &\cmark & &\cmark  & & & &\cmark & & & & \{$i_1$, $i_6$, $i_{8}$ \} & \{$i_1$, $i_5$, $i_9$ \}  & \{$i_1$, $i_3$, $i_7$ \}  \\ \hline
 ${\color[HTML]{FE0000} a_1}$  &{\color[HTML]{FE0000}\textbf{user 2}}  & & &  & &\cmark & & &\cmark & &  & \{$i_2$, $i_5$, $i_9$ \} & \{$i_2$, $i_5$, $i_7$ \}  & \{$i_1$, $i_5$, $i_8$ \}   \\ \hline
 ${\color[HTML]{FE0000} a_1}$  &{\color[HTML]{FE0000} \textbf{user 3}}   & & \cmark&  & & & &\cmark & & & & \{$i_1$, $i_6$, $i_7$\} & \{$i_2$, $i_5$, $i_9$ \} & \{$i_2$, $i_7$, $i_9$ \}  \\ \hline
 ${\color[HTML]{009901} a_2}$  &{\color[HTML]{009901} \textbf{user 4}}   & & &\cmark  &\cmark & & & & &\cmark & &  \{$i_3$, $i_4$, $i_9$\} & \{$i_4$, $i_5$, $i_6$ \} & \{$i_3$, $i_4$, $i_9$ \}   \\ \hline
 ${\color[HTML]{009901} a_2}$  &{\color[HTML]{009901} \textbf{user 5}}   & & &  & &\cmark & &\cmark & & &\cmark  &\{$i_1$, $i_5$, $i_{7}$\} & \{$i_1$, $i_2$, $i_{10}$ \}   &\{$i_5$, $i_7$, $i_{10}$ \}   \\ \hline
 ${\color[HTML]{009901} a_2}$  &{\color[HTML]{009901} \textbf{user 6}}  &\cmark & &\cmark  & & &\cmark & & &\cmark & &\{$i_2$, $i_{6}$, $i_{9}$\} & \{$i_1$, $i_5$, $i_8$ \} & \{$i_3$, $i_6$, $i_9$ \} \\ \hline
\end{tabular}
\vspace{-10pt}
\end{table*}

\begin{table}[tb]
\caption{Fairness of different recommenders in the toy example presented in Table \ref{tbl:toy_example} according to proposed GCE and individual-level accuracy metrics. Note that $p_{f_0}=[\frac{1}{2}, \frac{1}{2}]$ and $p_{f_1} = [\frac{2}{3}, \frac{1}{3}], p_{f_2} = [\frac{1}{3}, \frac{2}{3}]$ characterize the fair distribution as uniform or non-uniform distributions between
two groups.}
\label{tbl:fairness_results}
\vspace{-10pt}
\begin{tabular}{|c|c|c|c|c|c|c|}
\hline
      & \multicolumn{3}{c|}{\textbf{GCE} ($p_{f}$, $p$, $\alpha = -1$)} & \multirow{2}{*}{\textbf{P@3}}  & \multirow{2}{*}{\textbf{R@3}} \\ \cline{2-4}
      & $p_{f_0}$              & $p_{f_1}$& $p_{f_2}$             &  & \\ [2pt] \hline
\textbf{Rec 0} & 0.0800  &  0.3025    &0.0025 &$\frac{1}{2}$  &$\frac{1}{6}.\frac{19}{6} = 0.530$ \\ [2pt] \hline
\textbf{Rec 1} &  0       &  0.0625    &0.0625  & $\frac{1}{3}$  & $\frac{1}{6}.\frac{9}{4}=0.375$  \\ [2pt] \hline
\textbf{Rec 2} &  0.0078   &0.1182   &0.0244  & $1$  & $\frac{1}{6}.\frac{23}{4}=0.958$   \\[2pt] \hline
\end{tabular}
\vspace{-10pt}
\end{table}

\vspace{-7pt}
\section{Evaluating Fairness in RS}
In this section, we propose a framework based on generalized cross entropy for evaluating fairness in recommender systems. 
Let $\mathcal{U}$ and $\mathcal{I}$ denote a set of users and items, respectively. 
Suppose $\mathcal{A}$ be a set of sensitive attributes in which fairness is desired. 
Each attribute can be defined for either users, e.g., gender and race, or items, e.g., item provider (or stakeholder). 

The goal is to find an \emph{unfairness measure} $I$ that produces a non-negative real number for a recommender system. A recommender system $\mathcal{M}$ is considered less unfair (i.e., more fair) than $\mathcal{M}'$ with respect to the attribute $a \in \mathcal{A}$ if and only if $|I(\mathcal{M}, a)| < |I(\mathcal{M}', a)|$.
Previous works have used \emph{inequality} measures to evaluate algorithmic unfairness, however, we argue that fairness does not always imply equality. For instance, let us assume that there are two types of users in the system -- regular (free registration) and premium (paid) -- and the goal is to compute fairness with respect to the users' subscription type. In this example, it might be more fair to produce better recommendations for paid users, therefore, equality is not always equivalent to fairness. 
\ifmakeshorter
%
We define fairness of a recommender system as the \emph{generalized cross entropy} (GCE) for some parameter $\alpha \neq 
0 , 1$:
\fi
\begin{equation}
\label{eq:GCE}
    I(\mathcal{M}, x) = \frac{1}{\alpha (1-\alpha)}\left[\int p_f^\alpha(x) p^{(1-\alpha)}(x)~\diff x - 1\right]
\end{equation}
\ifmakeshorter
where $p$ and $p_f$ respectively denote the probability distribution of the system performance and the fair probability distribution, both with respect to the attribute $x = a$~\cite{botev2011generalized}.
\else
\fi
The unfairness measure $I$ is minimized with respect to attribute $x = a$ when $p = p_f$, meaning that the performance of the system is equal to the performance of a fair system. In the next sections, we discuss how to obtain or estimate these two probability distributions. 
%
If the attribute $a$ is discrete or categorical (as typical attributes, such as gender or race), then the unfairness measure is defined as:
\begin{equation}
\label{eq:GCE_discrete}
    I(\mathcal{M}, a) = \frac{1}{\alpha (1-\alpha)}\left[\sum_{a_j} p_f^\alpha(a_j) p^{(1-\alpha)}(a_j)  -1\right]
\end{equation}

\subsection{Fair Distribution $p_f$}
The definition of a fair distribution $p_f$ 
is problem-specific and should be determined based on the problem or target scenario in hand. For example, a job or music recommendation website may want to ensure that its premium users, who pay for their subscription, would receive more relevant recommendations. In this case, $p_f$ should be non-uniform across the user classes (premium versus free users). In other scenarios, a uniform definition of $p_f$ might be desired. Generally, when fairness is equivalent to equality, then $p_f$ should be uniform and in that case, the generalized cross entropy would be the same as generalized entropy (see \cite{speicher2018unified} for more information).


\subsection{Estimating Performance Distribution $p$}
The performance distribution $p$ should be estimated based on the output of the recommender system on a test set. In the following, we explain how we can compute this distribution for item attributes.  We define the recommendation gain ($rg_i$) for each item $i$ as follows:
\begin{equation}
\label{eq:rec_gain_item}
rg_i = \sum_{u \in \mathcal{U}} \phi(i,~ Rec^{K}_u) \ g(u, i, r)
\end{equation}
where $Rec^{K}_u$ is the set of top-$K$ items recommended by the system to the user $u \in U$. $\phi(i,~Rec^{K}_u) = 1$ if item $i$ is present in $Rec^{K}_u$; otherwise $\phi(i,~Rec^{k}_u) = 0$. The function $g(u,i,r)$ is the gain of recommending item $i$ to user $u$ with the rank $r$. Such gain function can be defined in different ways. In its simplest form, when $g(u,i,r) = 1$, the recommendation gain in Eq.~\ref{eq:rec_gain_item} would boil down to recommendation count (i.e., $rg_i = rc_i$). 
A binary gain in which $g(u,i,r) = 1$ when item $i$ recommended to user $u$ is relevant and $g(u,i,r) = 0$ otherwise, is another simple form of the gain function based on \textbf{relevance}.
The gain function $g$ can be also defined based on \textbf{ranking} information, 
i.e., recommending relevant items to users in higher ranks is given a higher gain. 
In such case, we recommend the discounted cumulative gain (DCG) function that is widely used in the definition of NDCG~\cite{jarvelin2002cumulated}, given by $\frac{2^{\text{rel}(u, i) - 1}}{\log_2 (r+1)}$ where $\text{rel}(u, i)$ denotes the relevance label for the user-item pair $u$ and $i$. We can further normalize the above formula based on the ideal DCG for user $u$ to compute the gain function $g$. 

The performance probability distribution $p$ is then proportional to the recommendation gain for the items associated to an attribute value $a_j$. Formally, the performance probability $p(a_j)$ used in Eq.~\eqref{eq:GCE_discrete} is computed as:
\ifmakeshorter
$p(a_j) = \sum_{i \in a_j} rg_i/Z $
\fi
where $Z$ is a normalization factor set equal to $Z=\sum_{i}{rg_i}$ to make sure that $\sum{p(a_j)} =1$.
Under an analogous formulation, we could define a variation of fairness for users based on Eq.~\eqref{eq:rec_gain_item}:
\begin{equation}
\label{eq:rec_gain_u}
rg_u = \sum_{i \in \mathcal{I}} \phi(i,~ Rec^{K}_u) \ g(u, i, r)
\end{equation}
where in this case, the gain function cannot be reduced to $1$, otherwise, all users would receive the same recommendation gain $rg_u$.

\section{Toy Example}
For the illustration of the proposed concept, in Table~\ref{tbl:toy_example} we provide a toy example on how our approach for fairness evaluation framework could be applied in a real recommendation setting. A set of six users belonging to two groups (each group is associated with an attribute value $a_1$ (red) or $a_2$ (green)) who are interacting with a set of items are shown in Table~\ref{tbl:toy_example}. Let us assume the red group represents users with a \textit{regular} (free registration) subscription type on an e-commerce website while the green group represents users with a \textit{premium} (paid) subscription type. A set of recommendations produced by different systems (\textbf{Rec0}, \textbf{Rec1}, and \textbf{Rec2}) are shown in the last columns. 
The goal is to compute fairness using the proposed fairness evaluation metric based on GCE given by 
Eq.~\eqref{eq:GCE_discrete}. 
The results of evaluation using three different evaluation metrics are shown in Table~\ref{tbl:fairness_results}. The metrics used for the evaluation of fairness and accuracy of the system include: (i) GCE (absolute value), (ii) Precision@3 and (iii) Recall@3.
Note that $GCE = 0$ means the system is completely fair, and the closer the value is to zero, the more fair the respective system is.

By looking at the recommendation results of \textbf{Rec0}, one can note that \textit{if fairness is defined in a uniform way between two groups}, defined through fair distribution $p_{f}$ = $[\frac{1}{2}, \frac{1}{2}]$, then \textbf{Rec0} is not a completely fair system, since $GCE = 0.08 \neq 0$. In contrast, \textit{if fairness is defined as providing recommendation of higher utility (usefulness) to green users who are users with paid premium membership type}, (e.g., by setting $p_{f}$ = $[\frac{1}{3}, \frac{2}{3}]$) then since $GCE \approx 0$, we can say that recommendations produced by \textbf{Rec0} are fair. Both of the above conclusions are drawn with respect to attribute \dquotes{subscription type} (with categories free/paid premium membership). This is an interesting insight which shows the evaluation framework is flexible enough to capture fairness based on the interest of system designer by defining what she considers as fair recommendation through the definition of $p_f$. While in many application scenarios we may define fairness as equality among different classes (e.g., gender, race), in some scenarios (such as those where the target attribute is not sensitive, e.g., regular v.s. premium users) fairness may not be equivalent to equality.

Furthermore, by comparing the performance results of \textbf{Rec1} and \textbf{Rec2}, we observe that, even though precision and recall improve for \textbf{Rec2} and becomes the most accurate recommendation list, it fails to keep a decent amount of fairness with respect to any parameter settings of GCE, 
as in both cases it is outperformed by the other methods. 
Moreover, GCE never reaches the optimal value, which in this case is attributed to the unequal distribution of resources among classes, since there are more relevant items on green than red users. 
This evidences that optimizing an algorithm to produce relevant recommendations does not necessarily result in more fair recommendation rather, conversely, a trade-off between the two evaluation properties can be noticed.

\vspace{-5pt}
\section{Experiments and Results}
In the section, we discuss our experimental setup and the results.
\vspace{-3pt}

\subsection{Data Descriptions}
We conduct experiments on two real-world datasets, Xing job recommendation dataset~\cite{Abel:2017:RCO:3109859.3109954} and Amazon Review dataset~\cite{Amazon:2019}. The datasets represent different item recommendation scenarios for job and e-commerce domains. We used Xing dataset to study the \textbf{item-related} notion of fairness, 
while Amazon is used to study the \textbf{user-related} notion of fairness. 

\noindent \textbf{Xing Job Recommendation Dataset (Xing-REC 17)}: The dataset was first released by XING as part of the ACM RecSys Challenge 2017 for a job recommendation task~\cite{Abel:2017:RCO:3109859.3109954}. 
The dataset contains 320M of interactions happened in over 3 months. 
The reason for choosing this dataset is that it provides several user-related attributes, such as \textit{membership types} (regular vs. premium), \textit{education degree}, and \textit{working country}, that can be useful for the study of fairness. For example, membership type allows us to study the non-equal (non-uniform) notion of fairness, as a recruiter may want to ensure premium users obtain better quality in their recommendations. 


\noindent \textbf{Amazon}: We used the toy and games subset which contains 53K preference scores by 1K users for 24K items, with a sparsity of $99.8\%$.
 We wanted the training set to be as close as possible to an on-line real scenario in which the recommender system is deployed, with this goal in mind we used a time-aware splitting.
 The most rigorous one would be the fixed-timestamp splitting method \cite{DBLP:journals/umuai/CamposDC14,DBLP:reference/sp/GunawardanaS15}.
 In these experiments, however, we adopted the methodology proposed in \cite{DBLP:conf/ecir/AnelliNSRT19} where a single timestamp is chosen, which represents the moment when test users are on the platform waiting for recommendations. The training set corresponds to the past interactions, and the performance is evaluated with data which correspond to future interactions. The splitting timestamp is selected to maximize the number of users involved in the evaluation according to two constraints: the training should retain at least 15 ratings, and the test set should contain at least 5 ratings.
 
    \vspace{-7pt}
\subsection{Experimental Setup}
\label{ss:exp}
Two recommendation scenarios are considered to evaluate the effectiveness of the proposed fairness evaluation framework with respect to \textbf{item-centric} or \textbf{user-centric} notion of fairness~\cite{burke2017multisided}.

\medskip
\noindent \textbf{Item fairness evaluation}: It applies the proposed fairness evaluation metrics based on GCE on the winner of the ACM RecSys Challenge 2017. The challenge was formulated as \dquotes{\textit{given a job posting, recommend a list of candidates that are suitable candidates for the job}}. As such, the user candidates are considered as target items for recommendation. In order to compute GCE, we used Eq.~\eqref{eq:rec_gain_item} by considering
a simplified case $g(u,i,r)=1$, in which the recommendation gain $rg_i$ boils down to recommendation count $rc_i$ for item $i$, i.e., the number of times each user appears in the recommendation lists of all jobs.

We compare two recommendation approaches: the winner submission and a random submission, and evaluate the systems' fairness from the perspective of users membership types, education, and location. As for membership type, premium users (or paid members) are expected to receive better quality of recommendation.

\begin{table}[tb]
\caption{Results of applying the proposed fairness evaluation metrics on Xing-REC 17 winner submission to identify \textit{item-centered} fairness for the attribute membership type (regular v.s. premium).  Note that in this case, it is desired to increase the utility of recommendation for premium (paid) users. $p_{f_0}=[1/2,1/2]$ (uniform) v.s. $p_{f_2}=[1/3,2/3]$ (non-uniform).}
\label{tbl:GCE_item1}
\vspace{-10pt}
\resizebox{\columnwidth}{!}{
\begin{tabular}{|c|c|c|c|c|c|}
\hline
  &\multicolumn{2}{c|}{\textbf{Membership type}}  & \multicolumn{2}{c|}{\textbf{GCE} ($p_{f}$, $p$, $\alpha = -1$)}   \\ \cline{2-5}
       &\textbf{regular}   &\textbf{premium} & $p_{f_0}$              & $p_{f_2}$            \\
       [2pt] \hline
\textbf{RSC Winner} & 4,108,771   &  547,029    &$0.2926$  &$0.6786$ \\ [2pt] \hline
\textbf{Random} & 4,209,878        & 445,759    & $0.3269$  & $0.7335$  \\ [2pt] \hline
\end{tabular}
}
\vspace{-10pt}
\end{table}

\begin{table}[tb]
\caption{Results of applying the proposed fairness evaluation metrics on Xing-REC 17 winner submission to identify \textit{item-centered} fairness. GCE1 and GCE2 have associated fair probability distributions equal to  $p_{f_0}$ = $[0.25,0.25,0.25,0.25]$, $p_{f_1}$ = $[0.7,0.1,0.1,0.1]$, $p_{f_2}$ = $[0.1,0.7,0.1,0.1]$, $p_{f_3}$ = $[0.1,0.1,0.7,0.1]$, $p_{f_4}$ = $[0.1,0.1,0.1,0.7]$ where $p_{f_0}$ defines fair distribution as uniform distribution while the rest define it as favoring each of groups}
\label{tbl:GCE_item2}
\vspace{-10pt}
\resizebox{\columnwidth}{!}{
\begin{tabular}{|l|l|l|l|l|l|l|l|l|l|}
\hline
 & \multicolumn{4}{c|}{\textbf{\begin{tabular}[c]{@{}c@{}} Country\end{tabular}}} &\multicolumn{5}{c|}{\textbf{\begin{tabular}[c]{@{}c@{}} GCE1 ($p_{f}$, $p$, $\alpha=-1)$ \end{tabular}}} \\ \cline{2-10} 
 & \textbf{German} & \textbf{Austrian} &\textbf{Swiss}  &\textbf{Other} &$p_{f_0}$ &$p_{f_1}$ &$p_{f_2}$ &$p_{f_3}$ &$p_{f_4}$ \\ \hline
winner & 3.9M & 156.1K &329.4K &186.4K &0.979  &0.061 &3.194 &3.177 & 3.192 \\ \hline
random & 3.8M & 253.6K &319.6K &239.8K &0.883 &0.038 & 2.945 &2.937 & 2.946 \\ \hline
 & \multicolumn{4}{c|}{\textbf{\begin{tabular}[c]{@{}c@{}} Education\end{tabular}}} &\multicolumn{5}{c|}{\textbf{\begin{tabular}[c]{@{}c@{}} GCE2 ($p_{f}$, $p$, $\alpha=-1)$ \end{tabular}}} \\ \cline{2-10} 
 & \textbf{NA} & \textbf{BSc} &\textbf{MSc}  &\textbf{PhD}  &$p_{f_0}$&$p_{f_1}$ &$p_{f_2}$ &$p_{f_3}$ &$p_{f_4}$\\ \hline
winner & 2.8M & 607.2K &1.0M &158.3K &0.398  &0.0974 &1.673 &1.547 & 1.741 \\ \hline
random & 3.0M & 428.4K &1.0M &203.4K &0.450 &0.0887 & 1.838 & 1.670& 1.866 \\ \hline
\end{tabular}
}
\vspace{-10pt}
\end{table}

\medskip

\noindent \textbf{User fairness evaluation}: Here, we experiment with the more traditional item recommendation task where we study the user fairness dimension.
We consider a scenario where a business owner may want to ensure superior recommendation quality for its more engaged users over less engaged (or new) users (or vice versa). 
In order to have a more intuitive sense about how fair different recommendation models are recommending to users of different classes, we study the fairness of different CF recommendation models with respect to users' interactions, defined in 4 categories: (i) very inactive (VIA), (ii) slightly inactive (SIA), (iii) slightly active (SA), and (iv) very active (VA). For each user, we compute the score $n_R(u)$ that corresponds to the total number of ratings provided by user $u$. 
We group the users in four groups according to the quartile that this score belongs to. 

\begin{table*}[tb]
\caption{Results of applying the proposed fairness evaluation metrics on Amazon dataset to identify user-centered fairness. 
The fair probability distributions are defined as $p_{f_i}$ so that $p_{f_i}(j)=0.1$ when $j\neq i$ and $0.7$ otherwise, except for $p_{f_0}$  that denotes the uniform distribution; i.e., just as in Table~\ref{tbl:GCE_item2}. 
Types of users (VIA/SIA/SA/VA) as defined in Section~\ref{ss:exp}.
Best values per column in bold.
}
\label{tbl:GCE_user}
\vspace{-10pt}
\small
\begin{tabular}{|c|c|c|c|c|c|c|c|c|c|c|c|}
\hline
 & \multicolumn{4}{c|}{\textbf{\begin{tabular}[c]{@{}c@{}} NDCG@10\end{tabular}}} &\multicolumn{5}{c|}{\textbf{\begin{tabular}[c]{@{}c@{}} GCE ($p_{f}$, $p$, $\alpha=-1)$ \end{tabular}}}  &\multicolumn{2}{c|}{\textbf{\begin{tabular}[c]{@{}c@{}} MAD  \end{tabular}}} \\ \cline{2-12}
 & \textbf{\begin{tabular}[c]{@{}l@{}}VIA\end{tabular}} & \textbf{\begin{tabular}[c]{@{}l@{}}SIA\end{tabular}} &\textbf{\begin{tabular}[c]{@{}l@{}}SA\end{tabular}} &\textbf{\begin{tabular}[c]{@{}l@{}}VA\end{tabular}} &$p_{f_0}$&$p_{f_1}$ &$p_{f_2}$ &$p_{f_3}$ &$p_{f_4}$&rating &ranking \\ \hline
Random&0.0000&0.0000&0.0000&0.0005&1.5000&4.5000&4.5000&4.5000&\textbf{0.2143}&\textbf{0.0000}&\textbf{0.0003}\\ \hline
MostPopular&0.0000&0.0006&0.0013&0.0014&0.2435&1.3586&1.2289&0.6714&0.5825&0.1864&0.0008\\ \hline
ItemKNN&0.0023&0.0021&0.0016&0.0036&0.0487&0.6218&0.6722&0.7537&0.2636&0.0254&0.0011\\ \hline
UserKNN&\textbf{0.0031}&\textbf{0.0040}&\textbf{0.0037}&\textbf{0.0053}&0.0214&0.6483&0.5379&0.5783&0.3319&0.0375&0.0012\\ \hline
BPRMF&0.0022&0.0025&0.0028&0.0016&\textbf{0.0191}&0.5496&\textbf{0.4767}&0.3881&0.6642&0.2078&0.0006\\ \hline
BPRSlim&0.0027&0.0023&0.0035&0.0017&0.0353&\textbf{0.5377}&0.6150&\textbf{0.3267}&0.7267&9.0009&0.0010\\ \hline
SVD++&0.0025&0.0025&0.0025&0.0042&0.0324&0.6336&0.6382&0.6361&0.2750&0.0027&0.0009\\ \hline
\end{tabular}
\vspace{-10pt}
\end{table*}


We have experimented with several recommendation models such as UserKNN~\cite{DBLP:conf/uai/BreeseHK98}, ItemKNN~\cite{sarwar2000analysis} (considering binarized and cosine similarity metric, Jaccard coefficient \cite{dong2011efficient}, and Pearson correlation \cite{DBLP:journals/ir/HerlockerKR02}),
SVD++~\cite{DBLP:conf/kdd/Koren08,koren2009matrix}, 
BPRMF~\cite{DBLP:conf/uai/RendleFGS09,Koren2009},
BPRSlim~\cite{DBLP:conf/icdm/NingK11}, and two non-personalized models: a most-popular algorithm and a random recommender.

For comparison with the proposed GCE metric, we include two complementary baseline metrics based on the absolute deviation between the mean ratings of different groups as defined in~\cite{DBLP:conf/cikm/ZhuHC18} $MAD(R^{(i)}, \ R^{(j)}) = \left | \frac{\sum{R^{(i)}}}{\left |R^{(i)}\right | }  - \frac{\sum{R^{(j)}}}{\left |R^{(j)} \right | }\right |$ where $R^{(i)}$ denotes the predicted ratings for all user-item combinations in group $i$ and $\left | R^{(i)} \right |$ is its size. Larger values for MAD mean larger differences between the groups, interpreted as unfairness. Given that our proposed GCE in user-fairness evaluation is based on NDCG, we adapt this definition to also compare between average NDCG for each group. 
We refer to these two baselines as \textbf{MAD-rating} and \textbf{MAD-ranking}. 
Finally, the reported MAD corresponds to the average MAD between all the pairwise combinations within the groups involved, i.e., $MAD = \mbox{avg}_{i,j}{(MAD(R^{(i)}, R^{(j)}))}$.
\vspace{-3mm}

\subsection{Results and Discussion}
We start our analysis with the results for the item fairness evaluation as described in Section~\ref{ss:exp}, presented in Tables~\ref{tbl:GCE_item1} and~\ref{tbl:GCE_item2}.
The counts in these tables represent the total number of users with a given category that each submission recommends. We observe in Table~\ref{tbl:GCE_item2} that recommendations produced by the RecSys Challenge winner performs better with $p_{f_0}$ than with $p_{f_2}$, since the GCE value is closer to $0$. This evidences that the proposed winner system produces balanced recommendations across the two membership classes. This is in contrast to our expectation that premium users should be provided better recommendations. Therefore, even though the winning submission could produce higher recommendation quality \emph{from a global perspective}, it does not comply with our expectation of a fair recommendation for this attribute, which is to recommend better recommendations to premium users. 


Furthermore, in Table~\ref{tbl:GCE_item2} we present the recommendation fairness evaluation results using GCE across two other attributes: Country and Education; each of these attributes takes $4$ categories. 
We define five variations of the fair distribution $p_f$: while $p_{f_0}$ considers all attribute categories equally important, the others give one attribute category a higher importance compared to the rest. After applying the GCE on the winner submission, we observe that with respect to the Country attribute, the lowest value of GCE (best case) is produced for the German companies ($GCE=0.061$) while for the Education attribute the category Unknown ($GCE=0.97$) produces the best outcome, in both cases, these categories are the most frequently recommended by the analyzed submission. 
These results show that for a given target application, if the system is looking for candidates with certain nationality (in this case, German) or education-level (here any), the system recommendations coming from the winner submission are closer to a fair system. 
In fact, due to the inherent biases in the dataset, the random submission is obtaining better results according to our definition of fairness for several of the fair distributions analyzed. However, it is worth mentioning that if the system designer wants to promote those users with BSc or PhD, the GCE would show that the winner submission provides better recommendations to those users than the random submission.
To the best of our knowledge, this is the first fairness-oriented evaluation metric that allows to capture these nuances, which as a consequence, helps on understanding how the recommendation algorithms work on each user group.

Now Table~\ref{tbl:GCE_user} shows the results for the proposed user fairness evaluation as described in Section~\ref{ss:exp}. 
We observe in this table that each recommender obtains a GCE value on a different range, an obvious consequence of the different performance obtained in each case for the different groups (as we observe in the NDCG@10 columns for each user type).
For instance, BPRMF is the one found by GCE to perform in a fair way when assuming uniformity with respect to the user groups ($p_{f_0}$), however, if the system designer aims to promote those recommenders that provide better suggestions to the most active users then Random, followed by ItemKNN and SVD++ are the most fair algorithms.

Comparing MAD against GCE, we observe that MAD-ranking produces lower results when NDCGs in each class are close to each other (e.g., in the case of Random recommender), which corresponds to the already discussed notion of fairness as equality/uniformity; similarly, MAD-rating obtains better results for the random algorithm because, as expected, such method has no inherent bias with respect to the defined user groups, but also for SVD++, probably because this recommender tends to predict ratings in a small range.
In both cases, it becomes evident that MAD, in contrast to our proposed GCE metric, cannot incorporate other definitions of fairness in its computation, hence, its flexibility is very limited.

In summary, we have shown that our proposed fairness evaluation metric is able to unveil whether a recommendation algorithm satisfies our definition of fairness, where we argue that it should emphasize a proper distribution of utility based on merits and needs.
We demonstrate this in both notions of fairness: based on users and based on items.
Therefore, we conclude that this metric could help better explaining the results of the algorithms towards specific groups of users and items, and as a consequence, it could increase the transparency of the recommender systems evaluation.

\section{Conclusion}
Fairness-aware recommendation research requires appropriate evaluation metrics to quantify fairness. 
Furthermore, fairness in RS can be associated with either items or users, even though this complementary view has been underrepresented in the literature. 
In this work, we have presented a probabilistic framework to measure fairness of RS under the perspective of users and items. Experimental results on two real-world datasets show the merits of the proposed evaluation framework.
In particular, one of the key aspects of our proposed evaluation metric is its transparency and flexibility, since it allows to incorporate domain knowledge (by means of an ideal fair distribution) that helps on understanding which item or user aspects the recommendation algorithms are over- or under-representing.

In the future, we plan to exploit the proposed fairness and relevance aware evaluation system to build recommender systems that directly optimize for this objective criterion. Also, it is of our interest to consider studying various fairness of recommendation under various content-based filtering or CF models using item content as side information~\cite{DBLP:journals/umuai/DeldjooDCECSIC19,DBLP:conf/recsys/DeldjooCEISC18} on different domains (e.g., tourism~\cite{adamczak2019session}, entertainment~\cite{DBLP:conf/iir/DeldjooSCP18}, social recommendation among others). Finally, we are considering to investigate the robustness of CF models against shilling attacks~\cite{deldjoo2019assess} crafted to undermine not only the accuracy of recommendations but also fairness of these models.  


\vspace{-4mm}

\section{Acknowledgements}
This work was supported in part by the Center for Intelligent Information Retrieval and in part by project TIN2016-80630-P (MINECO). 
Any opinions, findings and conclusions or recommendations expressed in this material are those of the authors and do not necessarily reflect those of the sponsors.

\bibliographystyle{ACM-Reference-Format}
\bibliography{recsys2019} 

\end{document}